\documentclass[5p, times, twocolumn, preprint, sort&compress]{elsarticle}

\usepackage{hyperref}
\usepackage{changepage}
\usepackage{listings}
\usepackage{xcolor}
\usepackage{adjustbox}
\usepackage{amsmath}
\usepackage{array}
\usepackage{booktabs}
\usepackage{graphicx}
\usepackage{subfig}
\usepackage[none]{hyphenat}
\usepackage{algorithm}
\usepackage{algorithmic}
\usepackage[perpage]{footmisc}
\usepackage{stfloats}

\definecolor{codegreen}{rgb}{0,0.6,0}
\definecolor{codegray}{rgb}{0.5,0.5,0.5}
\definecolor{codepurple}{rgb}{0.58,0,0.82}
\definecolor{codeyellow}{rgb}{0.51, 0.37, 0.012}

\lstdefinelanguage{python} {
	keywords={and, as, assert, break, class, continue, def, del, elif, else, except, False, finally, for, from, global, if, import, in, is, lambda, None, nonlocal, not, or, pass, raise, return, True, try, while, with, yield},
	ndkeywords={},
	sensitive=false,
	comment=[l]{\#},
	morecomment=[s]{'''}{'''},
	morecomment=[s]{"""}{"""},
	morestring=[b]',
	morestring=[b]"
}

\begin{document}
	\begin{frontmatter}
		
		\title{A Multi-Objective Framework for Optimizing GPU-Enabled VM Placement in Cloud Data Centers with Multi-Instance GPU Technology}
		
		\author[aff]{Ahmad Siavashi}
		\ead{siavashi@aut.ac.ir}
		
		\author[aff]{Mahmoud Momtazpour}
		\ead{momtazpour@aut.ac.ir}
		
		\address[aff]{Department of Computer Engineering, Amirkabir University of Technology}
		
		
			\begin{abstract}
	The extensive use of GPUs in cloud computing and the growing need for multitenancy have driven the development of innovative solutions for efficient GPU resource management. Multi-Instance GPU (MIG) technology from NVIDIA enables shared GPU usage in cloud data centers by providing isolated instances. However, MIG placement rules often lead to fragmentation and suboptimal resource utilization. In this work, we formally model the MIG-enabled VM placement as a multi-objective Integer Linear Programming (ILP) problem aimed at maximizing request acceptance, minimizing active hardware usage, and reducing migration overhead. Building upon this formulation, we propose GRMU, a multi-stage placement framework designed to address MIG placement challenges. GRMU performs intra-GPU migrations for defragmentation of a single GPU and inter-GPU migrations for consolidation and resource efficiency. It also employs a quota-based partitioning approach to allocate GPUs into two distinct baskets: one for large-profile workloads and another for smaller-profile workloads. Each basket has predefined capacity limits, ensuring fair resource distribution and preventing large-profile workloads from monopolizing resources. Evaluations on a real-world Alibaba GPU cluster trace reveal that GRMU improves acceptance rates by 22\%, reduces active hardware by 17\%, and incurs migration for only 1\% of MIG-enabled VMs, demonstrating its effectiveness in minimizing fragmentation and improving resource utilization.
		\end{abstract}
		
		
	\end{frontmatter}
	
	\section{Introduction}
	\label{s:introduction}
	
	The rapid expansion of machine learning, data analytics, and high-performance computing (HPC) workloads has positioned GPUs at the heart of cloud computing services. While GPUs offer exceptional parallel processing capabilities, they also introduce challenges in scheduling and resource sharing. Traditional scheduling methods, designed for standalone, monolithic GPUs, struggle to efficiently allocate resources when multiple tenants need to share a single GPU \cite{siavashi2023gvmp,chung_fine-grained_2025,amaral2017,zhu2021,li2022miso}.
	
	NVIDIA’s MIG technology partially addresses this challenge by partitioning a single GPU into multiple isolated instances, each with dedicated memory and compute engines. This approach enhances performance isolation and utilization, enabling multiple workloads to run on the same GPU. However, MIG enforces rigid rules for block alignment and profile placement, which can lead to fragmentation—where memory blocks become unusable or inefficiently allocated, ultimately reducing overall resource efficiency \cite{lee2024,weng_beware_2023}.
	
	In large-scale cloud environments, fragmentation worsens as MIG partitions are continuously created and removed to accommodate shifting user demands. A naive or short-sighted allocation strategy can quickly lock GPUs into a suboptimal state, limiting the acceptance of high-priority requests and increasing the number of active GPUs. To our knowledge, no existing placement strategy fully accounts for MIG’s strict placement constraints during allocation.
	
	This article introduces a multi-objective ILP model that represent MIG-enabled VM allocation constraints across a multi-GPU, multi-host data center. Our goal is to: (i) maximize the acceptance rate of incoming workloads, (ii) minimize the number of active hardware, and (iii) reduce migration overhead. We also present GRMU, a framework that integrates quota-based partitioning, defragmentation, and consolidation strategies. GRMU categorizes GPUs into heavy and light baskets to accommodate large profiles (e.g., 7g.40gb) while preserving space for smaller profiles. Additionally, it employs migrations to reclaim unused blocks and consolidate workloads on partially utilized GPUs, optimizing overall resource efficiency.
	
	By evaluating GRMU on real-world traces from the Alibaba GPU cluster \cite{weng_beware_2023,noauthor_clusterdatacluster-trace-gpu-v2023_nodate,siavashi_gpu_2024}, we illustrate that carefully tuning basket capacities and consolidation intervals yields substantial improvements in acceptance rates and resource utilization while keeping migration overhead at a minimum. Our results showcase a 22\% higher acceptance rate, a 17\% reduction in active hardware, and only 1\% of accepted requests undergoing migration—an appealing trade-off for cloud providers aiming to optimize performance and hardware usage.
	
	The remainder of this paper is structured as follows. Section \ref{s:related-work} reviews related work, and Section \ref{s:background} introduces MIG technology and its placement rules. Section \ref{s:motivation} highlights placement challenges, while Section \ref{s:default-placement-policy} describes built-in MIG placement strategy. Section \ref{s:formulation} formalizes the problem as an ILP model, and Section \ref{s:methodology} presents our proposed VM placement solution, GRMU. Section \ref{s:evaluation} provides an evaluation of GRMU, and Section \ref{s:conclusion} concludes with future directions.
	
	\section{Related Work}
	\label{s:related-work}
	
	The problem of VM placement in cloud data centers is well studied \cite{dias2021,imran2022,saidi2023task,lin2024energy,alahmad2024multiple,xu2024meta,regaieg2021multi}. GPU-enabled VMs introduce additional challenges due to GPU virtualization constraints \cite{hong2017gpu}. NVIDIA introduced GRID \cite{nvidia_grid_doc}, which supported only homogeneous tenants. Later, MIG \cite{nvidia_mig_doc} enabled both homogeneous and heterogeneous instances. In \cite{siavashi2019gpucloudsim}, NVIDIA GRID is examined for GPU-enabled VM placement, highlighting inefficiencies in the first-fit strategy as VM numbers grow. The proposed first-fit increasing algorithm improves acceptance rate (59\%), makespan (25\%), and energy consumption (21\%). In \cite{kulkarni2021gpu}, NVIDIA GRID is further analyzed for GPU-aware VM placement, enhancing energy efficiency, reducing host search space, and shortening makespan while tackling GPU resource management challenges in cloud applications.
	
	An ILP model for GPU-enabled VM placement using NVIDIA GRID GPU virtualization is proposed in \cite{garg2019virtual}, aiming to minimize physical GPU usage in the cloud. It is compared with two heuristics, showing similar performance to ILP while achieving faster execution. A metaheuristic approach using dense neural networks (DNN) was proposed in \cite{sivaraman2019tecn}. The authors built a simulator for GPU-enabled VM placement with NVIDIA GRID, implemented six heuristics, and trained DNNs on various configurations. They also explored combining multiple DNNs. Using a custom metric based on GPU utilization and job waiting time, their method outperformed heuristics in 76\% of test cases. In \cite{sivaraman2018task}, authors explore a GPU-enabled VM placement problem for NVIDIA GRID, optimizing performance using a cost function based on GPU utilization, fully executed VMs, and GPU execution time. Various VM placement policies are analyzed, with one achieving the best results. 
	
	In \cite{li2022miso}, authors highlight the rapid advancements in GPU technology, enhancing HPC and AI/ML research but also leading to inefficient resource utilization. To address this, they propose MISO, a technique leveraging MIG on NVIDIA datacenter GPUs (e.g., A100, H100) for dynamic GPU resource partitioning. MISO utilizes the lightweight Multi-Process Service (MPS) \cite{nvidia_mps_doc} to predict optimal MIG partitioning without incurring exploration overhead. This approach improves GPU efficiency, achieving 49\% and 16\% lower average job completion times compared to unpartitioned and optimal static GPU partitioning, respectively.
	
	To optimize MIG partitions for serving DNNs, \cite{tan2021serving} introduces the Reconfigurable Machine Scheduling (RMS) problem, a new NP-hard abstraction. They propose MIG-serving, an algorithmic pipeline using greedy heuristics, Genetic Algorithm (GA), and Monte Carlo Tree Search (MCTS) for cost-efficient deployments. MIG-serving includes an optimizer for deployment generation and a controller for seamless application. Experiments on a 24-GPU A100 cluster show a 40\% GPU reduction over baseline static allocations with minimal service disruptions. In \cite{lee2024parvagpu}, the authors propose ParvaGPU, a GPU space-sharing framework that integrates MIG and MPS to optimize DNN inference in cloud environments. It mitigates GPU underutilization with the Optimal Triplet Decision algorithm, which selects MPS-activated MIG instances for maximum throughput. The Demand Matching algorithm optimally assigns instances to meet high request rates. To reduce GPU non-contiguous space inefficiency, Segment Relocation redistributes instances, while Allocation Optimization minimizes fragmentation by resizing partitions. Evaluations on A100 GPUs show that ParvaGPU ensures Service-Level Objective (SLO) compliance while reducing GPU usage.
	
	In \cite{arima2022optimizing}, the authors integrate GPU partitioning with power-aware scheduling to optimize resource allocation in CPU-GPU heterogeneous systems under power constraints. Leveraging MIG for fine-grained workload co-location, their approach employs scalability and interference models to improve efficiency and achieve near-optimal configurations across diverse workloads. In \cite{saroliya2023hierarchical}, authors integrate MPS and MIG hierarchically, using reinforcement learning to jointly optimize partitioning and job co-scheduling. Experimental results demonstrate up to 1.87× throughput improvement over time-sharing scheduling.
	
	In \cite{weng_beware_2023}, authors identify GPU fragmentation as a key inefficiency in large ML clusters, where traditional bin packing fails to address partial GPU allocations. They propose Fragmentation Gradient Descent (FGD), which quantifies fragmentation and schedules tasks along its steepest descent to optimize GPU use. Evaluated on a 6,200-GPU cluster, FGD reduces unallocated GPUs by 49\%, reclaiming 290 more GPUs than existing schedulers.
	
	Our work differs from previous studies by considering MIG constraints, which impose unique placement rules. We formulate the problem such that MIG-enabled VMs require entire, indivisible GPU instances. Additionally, we optimize across three conflicting objectives: maximizing acceptance, minimizing active hardware, and reducing migrations. Furthermore, we enforce a stricter definition of active hardware, counting idle GPUs only when the entire machine is idle.
	
	\section{Background}
	\label{s:background}
	
	NVIDIA Ampere and Hopper GPUs feature MIG technology \cite{nvidia_mig_doc}, which splits a single GPU into multiple isolated and heterogeneous instances, each with their own compute engines and memory. Performance isolation aims to prevent interference between workloads on a single GPU, ensuring that each workload can run independently without affecting the others. Important for high-density computing, MIG enables effective concurrent GPU use by multiple users and applications.
	
	The NVIDIA A100, a MIG-enabled GPU, comprises 7 compute engines and 8 memory blocks, each providing 5 GB of memory for a total capacity of 40 GB. It supports six distinct MIG GPU Instance (GI) profiles, as shown in Table \ref{tab:a100-gis}, with each profile combining specific numbers of compute engines (\( C \)) and memory blocks (\( M \)). These profiles follow the naming convention \( Cg.Mgb \). While enabling MIG mode is a necessary first step, creating GPU instances and their corresponding compute instances (CIs) is also required. However, this paper focuses exclusively on GIs, leaving details about CIs outside its scope.
	
	\begin{table}[h]
		\centering
		\scriptsize
		\caption{GPU instance profiles on A100}
		\begin{tabular}{|c|c|c|c|}
			\hline
			\textbf{Profile Name} & \textbf{Memory Fraction} & \textbf{Compute Engines} & \textbf{Instances Available} \\ \hline
			MIG 1g.5gb            & 1/8                         & 1/7                                  & 7                                     \\ \hline
			MIG 1g.10gb           & 2/8                         & 1/7                                  & 4                                     \\ \hline
			MIG 2g.10gb           & 2/8                         & 2/7                                  & 3                                     \\ \hline
			MIG 3g.20gb           & 4/8                         & 3/7                                  & 2                                     \\ \hline
			MIG 4g.20gb           & 4/8                         & 4/7                                  & 1                                     \\ \hline
			MIG 7g.40gb           & 8/8                         & 7/7                                  & 1                                     \\ \hline
		\end{tabular}
		\label{tab:a100-gis}
	\end{table}

\begin{figure}[h]
		\centering
		\includegraphics[trim=0.7cm 0.3cm 0.7cm 0.3cm, clip,, width=1\columnwidth]{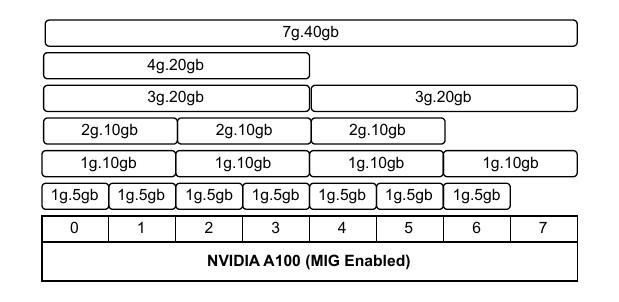}
		\caption{Profile placements on A100}
		\label{fig:a100-pp}
	\end{figure}

	Specific rules determine the allocation of GIs on MIG-enabled A100 GPUs, as detailed in NVIDIA's documentation \cite{nvidia_mig_doc}. This work adopts a perspective focused on memory blocks rather than compute engines for greater intuitiveness. Fig.~\ref{fig:a100-pp} shows the GPU layout and starting memory blocks for GIs, with 78 valid combinations formed by aligning profiles from left to right without vertical overlaps. These allocation principles are followed in other MIG-enabled GPUs.
	
	\section{Motivation}
	\label{s:motivation}
	
\begin{figure}[h]
		\centering
		\subfloat[Non-contiguous memory blocks]{
			\includegraphics[trim=0.7cm 0.3cm 0.7cm 0.3cm, clip,,width=0.45\columnwidth]{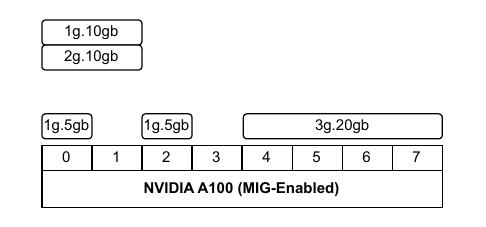}
			\label{fig:frag1}
		}
		\hfill
		\subfloat[Restriction on start block]{
			\includegraphics[trim=0.7cm 0.3cm 0.7cm 0.3cm, clip,,width=0.45\columnwidth]{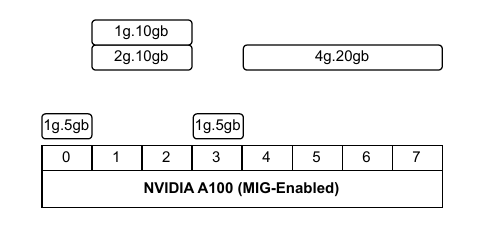}
			\label{fig:frag2}
		}\\
		\subfloat[Enhanced utilization through GPU block rearrangement]{
			\includegraphics[trim=0.7cm 0.1cm 0.7cm 0.1cm, clip,,width=0.97\columnwidth]{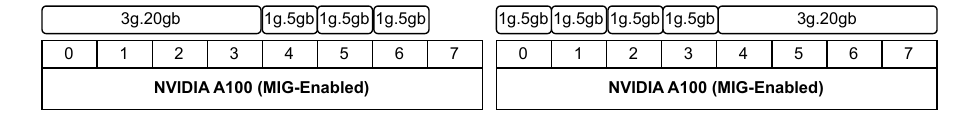}
			\label{fig:frag3}
		}
		\caption{MIG fragmentation scenarios}
		\label{fig:a100-frag}
	\end{figure}

	The rules for placing MIG profiles present challenges in the allocation of GIs. GIs are created and destroyed in response to accepted and completed requests, which can lead to fragmentation. Fig.~\ref{fig:a100-frag}(a) depicts a common fragmentation scenario where, due to non-contiguous memory blocks, it is impossible to allocate either 1g.10gb or 2g.10gb GIs. In MIG, however, having contiguous free memory blocks does not always ensure GI allocation. Fig.~\ref{fig:a100-frag}(b) depicts a situation where the displayed MIG profiles cannot be assigned to the specified contiguous blocks, owing to the lack of availability of their necessary starting blocks. Furthermore, 
	Fig.~\ref{fig:a100-frag}(c) illustrates a scenario where rearranging the allocated blocks on the GPU allows for the assignment of additional profiles, thus enhancing the utilization of the device. These illustrations show traditional and MIG-specific fragmentation scenarios.
	
	Tackling defragmentation challenges in data centers with numerous GPUs is essential, as these problems intensify in such environments. The difficulty in allocating GIs on single GPUs, due to non-contiguous memory and unavailable starting blocks, limits utilization. Reorganizing GPU blocks to add more profiles can improve device efficiency. Tackling these problems can enable providers to handle more requests effectively.
	
	\section{Default Placement Policy}
	\label{s:default-placement-policy}
	
	To quantify the versatility of a GPU configuration \( G \) in accommodating various profiles, we introduce the metric Configuration Capability (\( CC \))\footnote{Distinct from CUDA Compute Capability}. Mathematically, assuming \( \mathcal{P} \) represents the set of all supported profiles and \( S(G, p) \) indicates the available initial blocks for profile \( p \) within configuration \( G \), \( CC \) is defined as:
	
	\begin{align}
		\label{eq:cc}
		CC = \sum_{p \in \mathcal{P}} |S(G, p)|
	\end{align}	
	
	For example, the configuration $G = \{1,2,4,5,6,7\}$ in \text{Fig.~\ref{fig:a100-frag}(b)} has a \( CC = 9 \) because it can accommodate \( 5 \times 1g.5\text{gb} \), \( 2 \times 1g.10\text{gb} \), \( 1 \times 2g.10\text{gb} \), and \( 1 \times 3g.20\text{gb} \) profiles. The default MIG profile placement policy observed in our experiments with NVIDIA driver \( 530.30.02 \) positions a profile within the given configuration to maximize the \( CC \) value. For a given GI profile \( p_{req} \), the policy selects the starting block \( Y \) within the GPU that maximizes \( CC \), as represented mathematically below:
	
	\begin{align}
		\label{eq:max-cc}
		\underset{Y \in S(G, p_{req})}{\text{arg max}} \sum_{p_{alt} \in \mathcal{P}} |S(G \setminus Y, P_{alt})|
	\end{align} 
	
	This placement approach is further detailed in Algorithm \ref{alg:cc}. In lines 1 to 8, the start blocks for each MIG profile are defined. The function \textsc{GetCC} takes a GPU configuration and the list of all profiles with their start blocks. For each profile, it counts the number of start blocks where the profile can be placed and returns the total count, which corresponds to the \( CC \) value. The function \textsc{Assign} places a given \texttt{profile} on a GPU configuration \texttt{gpu} to maximize the \( CC \) value. It tests placing the profile at all possible start positions, calculates the \( CC \) value for each resulting configuration, and finally assigns the profile to the blocks that yield the highest \( CC \) value. Table \ref{tab:functions} lists the helper functions used in all pseudocodes in this paper.
	
	\begin{algorithm}[h]
		\caption{\small GPU Blocks Allocation}
		\footnotesize
		\label{alg:cc}
		\begin{algorithmic}[1]
			
			\STATE \texttt{startBlocks} $=$ \{
			\STATE \quad \texttt{1g.5gb}: $\{0, 1, 2, 3, 4, 5, 6\}$,
			\STATE \quad \texttt{1g.10gb}: $\{0, 2, 4, 6\}$,
			\STATE \quad \texttt{2g.10gb}: $\{0, 2, 4\}$,
			\STATE \quad \texttt{3g.20gb}: $\{0, 4\}$,
			\STATE \quad \texttt{4g.20gb}: $\{0\}$,
			\STATE \quad \texttt{7g.40gb}: $\{0\}$
			\STATE \}
			
			\vspace{0.1cm}
			
			\STATE \textbf{Function} \textsc{GetCC}\texttt{(G)}
			\STATE \quad \texttt{CC} $\leftarrow 0$
			
			\STATE \quad \textbf{for} \texttt{profile, starts} $\in$ \texttt{startBlocks} \textbf{do}
			\STATE \quad\quad \textbf{for} \texttt{start} $\in$ \texttt{starts} \textbf{do}
			\STATE \quad\quad\quad \texttt{blocks} $\leftarrow \{ \texttt{start} + i \mid i < \textsc{Size}\texttt{(profile)} \}$
			\STATE \quad\quad\quad \textbf{if} \texttt{blocks} $\subseteq$ \texttt{G} \textbf{then}
			\STATE \quad\quad\quad\quad \texttt{CC} $\leftarrow$ \texttt{CC} $+ 1$
			
			\STATE \quad \textbf{return} \texttt{CC}
			
			\vspace{0.1cm}
			
			\STATE \textbf{Function} \textsc{Assign}\texttt{(profile, gpu)}
			\STATE \quad \texttt{bestBlocks} $\leftarrow \emptyset$
			\STATE \quad \texttt{maxCC} $\leftarrow -1$
			
			\STATE \quad \textbf{for} \texttt{start} $\in$ \texttt{startBlocks[profile]} \textbf{do}
			\STATE \quad\quad \texttt{blocks} $\leftarrow \{ \texttt{start} + i \mid i < \textsc{Size}\texttt{(profile)} \}$
			\STATE \quad\quad \textbf{if} \texttt{blocks} $\subseteq$ \texttt{gpu} \textbf{then}
			\STATE \quad\quad\quad \texttt{CC} $\leftarrow$ \textsc{GetCC}\texttt{(gpu} $\setminus$ \texttt{blocks})
			\STATE \quad\quad\quad \textbf{if} \texttt{CC} $>$ \texttt{maxCC} \textbf{then}
			\STATE \quad\quad\quad\quad \texttt{bestBlocks} $\leftarrow$ \texttt{blocks}
			\STATE \quad\quad\quad\quad \texttt{maxCC} $\leftarrow$ \texttt{CC}
			
			\STATE \quad \texttt{gpu} $\leftarrow$ \texttt{gpu} $\setminus$ \texttt{bestBlocks} \quad \texttt{// Allocate profile on GPU}
			\STATE \quad \textbf{return} \textbf{True} \textbf{if} \texttt{bestBlocks} \textbf{else} \textbf{False}
			
		\end{algorithmic}
	\end{algorithm}

\begin{table}[h]
	\centering
	\scriptsize
	\caption{Definitions of helper functions in pseudocodes}
	\begin{tabular}{|l|l|}
		\hline
		\textbf{Function} & \textbf{Description} \\
		\hline
		\textsc{Size}(\texttt{p}) & Memory blocks of MIG profile \\
		\textsc{Add}(\texttt{l, g}) & Adds GPU \texttt{g} to list \texttt{l} by \texttt{globalIndex} \\
		\textsc{Get}(\texttt{l}) & Pops first GPU from list \texttt{l} \\
		\textsc{Remove}(\texttt{l, g}) & Removes GPU \texttt{g} from list \texttt{l} \\
		\textsc{Copy}(\texttt{g}) & Clones GPU \texttt{g} \\
		\textsc{HalfFull}(\texttt{g}) & \texttt{true} if lower or upper half of \texttt{g} is occupied \\
		\textsc{SingleProfile}(\texttt{g}) & \texttt{true} if \texttt{g} has only one profile \\
		\textsc{Max}(\texttt{l, f}) & Returns element in \texttt{l} with max \texttt{f} value \\
		\textsc{Relocated}(\texttt{g, g2}) & Returns VMs positioned differently in \texttt{g} and \texttt{g2} \\
		\textsc{InterMigrate}(\texttt{s, d}) & Moves guest VMs from \texttt{s} GPU to \texttt{d} GPU \\
		\textsc{IntraMigrate}(\texttt{v, g}) & Relocates VMs in \texttt{v} on \texttt{g} \\
		\hline
	\end{tabular}
	\label{tab:functions}
\end{table}

	\subsection{Configurations Analysis}
	\label{ss:analysis}
	
	We employ depth-first search to develop a tree that consists of all possible memory configurations for a single A100 GPU, starting from a state where the GPU is not in use. We generate a new child node for each configuration by adding a new GI. This addition of GIs continues until the GPU cannot accommodate any more, resulting in 78 terminal nodes where no further GIs can be added. The finalized tree encompasses 723 unique configurations. In practical scenarios, the addition or removal of GIs shifts the configuration of the GPU to one of these identified configurations.
	
	Configurations may have the same GIs but show different $CC$ values based on how the GIs are arranged in their blocks. Defining an optimal configuration as one that attains the highest $CC$ value among all arrangements with the same set of GIs, it is found that 67\% of the 723 configurations, or 482 in total, are in suboptimal arrangements. In multi-GPU environments, the probability of encountering different GI profiles becomes more crucial. 
	
	The default placement policy, which sequentially assigns GIs, results in 248 configurations—representing 34\% of the total 723 possible configurations in the space—excluding those influenced by system dynamics, such as GI departures. This sequential assignment approach can lead to suboptimal GPU states. Take, for instance, the arrival of two 1g.5gb GI profiles. Ideally, in an empty GPU, these GIs should occupy memory blocks 4 and 5. However, treating GIs independently may lead to them being placed in blocks 4 and 6, which is not the most efficient allocation. Approximately 69\%, equating to 172 out of the 248 allocations made by the default policy, fall short of being optimal. Moreover, since the policy does not reposition GIs post-allocation, a configuration that was initially optimal can become suboptimal following the departure and subsequent arrival of GIs.
	
\begin{figure}[h]
		\centering
		\subfloat[Original configuration]{
			\includegraphics[trim=0.7cm 0.1cm 0.7cm 0.1cm, clip,,width=0.48\columnwidth]{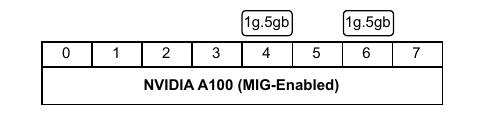}
			\label{fig:frag1}
		}
		\hfill
		\subfloat[Alternative configuration]{
			\includegraphics[trim=0.7cm 0.1cm 0.7cm 0.1cm, clip,,width=0.48\columnwidth]{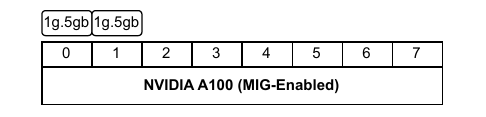}
			\label{fig:frag2}
		}\\
		\caption{Alternative configurations with different per profile capacity}
		\label{fig:a100-alt-configs}
	\end{figure}

	The $CC$ metric effectively indicates a configuration's capacity to accommodate new GIs, but it simplifies the optimality calculation by treating different profiles as equivalent. This approach neglects real-world conditions where certain GIs, such as 1g.10gb, might be more likely to occur. Contrary to what might be assumed, sufficient space for a larger profile does not guarantee the fit of smaller ones due to each profile's specific block fitting requirements. Fig.~\ref{fig:a100-alt-configs} shows an alternative to a configuration determined by the default allocation policy. As highlighted in Table \ref{tab:alt-configs}, despite possessing the same $CC$, the alternative configuration accommodates more 1g.10gb profiles, albeit at the expense of one fewer 4g.20gb profile.
	
	\begin{table}[h]
		\centering
		\scriptsize
		\begin{tabular}{l|c|c}
			\textbf{Profile} & \textbf{Original Configuration} & \textbf{Alternative Configuration} \\
			\hline
			1g.5gb  & 5 & 5 \\
			1g.10gb & 2 & 3 \\
			2g.10gb & 2 & 2 \\
			3g.20gb & 1 & 1 \\
			4g.20gb & 1 & 0 \\
			7g.40gb & 0 & 0 \\
			\hline
			$CC$ & 11 & 11 \\
			\hline
		\end{tabular}
		\caption{Per profile capacity of original and alternative configurations}
		\label{tab:alt-configs}
	\end{table}
	
	In a single GPU with 723 unique configurations, 19\% or 138 configurations can support a specific profile more effectively than alternative configurations that have the same profiles arranged in different blocks, despite these alternatives having the same or lower $CC$ value. With two GPUs, there are $\binom{723 + 2 - 1}{2} = 261,726$ distinct configurations and in 79\%, or 205,575 cases, alternative configurations exist that can accommodate at least one type of profile more effectively, even if they have the same or lower $CC$ compared to the original arrangement. With the addition of more GPUs, the complexity of the configuration space grows significantly, making a thorough examination increasingly difficult, or perhaps impractical.
	
	These analyses offer readers key insights into placement intricacies, as a thorough understanding of MIG profile placement is essential for effective resource allocation. Given that NVIDIA does not permit modifications to MIG placement policies within a single GPU, this work develops a VM placement policy for MIG-enabled cloud data centers, adhering to the GPU native allocation mechanisms.
	
	\section{Problem Definition and Objectives}
	\label{s:formulation}
	
	The problem is modeled as an online stochastic process using ILP, wherein requests arrive gradually, each with requirements that are unknown a priori, and placement decisions include the capability for preemption. We assume a discrete-time approach in which each discrete interval is used for evaluating incoming requests and making placement decisions. Time indices have been excluded to enhance readability. The model supports heterogeneity in workloads and infrastructure configurations, closely mirroring real-world environments.
	
	Consider $N$ as the set of virtual machines (VMs) and $M$ as the set of physical machines (PMs). We refer to the $i^{th}$ VM in $N$ as $\text{VM}_i$ and the $j^{th}$ PM in $M$ as $\text{PM}_j$. The collection of GPUs in $\text{PM}_j$ is represented by $P_j$. The variable $x_{ij}$ is used to indicate the allocation of $\text{VM}_i$ to $\text{PM}_j$. Similarly, $y_{ijk}$ represents the assignment of the GI of $\text{VM}_i$ to the $k^{th}$ GPU of $\text{PM}_j$, and $z_{ijk}$ signifies the starting offset of the GI of $\text{VM}_j$ in the $k^{th}$ GPU of $\text{PM}_j$. The essential symbols are provided in Table \ref{tab:notations}. Using these definitions, the problem is formulated as follows:
	
	\begin{align}
		\begin{split}
			\label{eq:1}
			& \textbf{Maximize} \quad  \sum_{i \in N} \sum_{j \in M} a_{i} x_{ij}
		\end{split}\\
		\begin{split}
			\label{eq:2}
			& \textbf{Minimize} \quad  \sum_{j \in M} b_j (\varphi_{j} + \sum_{k \in P_j} \gamma_{jk})
		\end{split}\\
		\begin{split}
			\label{eq:3}
			& \textbf{Minimize} \quad  \sum_{i \in N} \sum_{j \in M} \delta_{i} (m_{ij} + \sum_{k \in P_j}\omega_{ijk})
		\end{split}\\
		\notag	& \textbf{Subject to} \\
		\begin{split}
			\label{eq:4}
			\sum_{i \in N} x_{ij} c_{i} \leq C_{j} \quad \forall j \in M 
		\end{split}\\
		\begin{split}
			\label{eq:5}
			\sum_{i \in N} x_{ij} r_{i} \leq R_{j} \quad \forall j \in M 
		\end{split}\\
		\begin{split}
			\label{eq:6}
			\sum_{j \in M} x_{ij} \leq 1 \quad \forall i \in N
		\end{split}\\
		\begin{split}
			\label{eq:7}
			\sum_{j \in M} \sum_{k \in P_j} y_{ijk} \leq 1 \quad \forall i \in N
		\end{split}
	\end{align}
	\begin{align}
		\begin{split}
			\label{eq:8}
			x_{ij} \leq \sum_{k \in P_j} y_{ijk} \quad \forall i \in N, \forall j \in M \\
		\end{split}\\
		\begin{split}
			\label{eq:9}	
			y_{ijk} \leq x_{ij} \quad \forall i \in N, \, \forall j \in M, \, \forall k \in P_j
		\end{split}\\
		\begin{split}
			\label{eq:10}
			z_{ijk} + g_i y_{ijk} \leq z_{i'jk} + B \alpha_{ii'jk} \quad \forall i,i' \in N, i \neq i', \forall j \in M, \forall k \in P_j
		\end{split}\\
		\begin{split}
			\label{eq:11}
			z_{i'jk} + g_{i'} y_{i'jk} \leq z_{ijk} + B (1 - \alpha_{ii'jk}) \quad \forall i,i' \in N, i \neq i', \forall j \in M, \forall k \in P_j\\
		\end{split}\\
		\begin{split}
			\label{eq:12}
			z_{ijk} \leq g_i \beta_i + B (1 - y_{ijk}) \quad \forall i \in N, \forall j \in M, \forall k \in P_j\\
		\end{split}\\
		\begin{split}
			\label{eq:13}
			-z_{ijk} \leq -g_i \beta_i + B (1 - y_{ijk}) \quad \forall i \in N, \forall j \in M, \forall k \in P_j\\
		\end{split}\\
		\begin{split}
			\label{eq:14}
			z_{ijk} \leq s_i \quad \forall i \in N, \forall j \in M, \forall k \in P_j\\
		\end{split}\\
		\begin{split}
			\label{eq:15}
			h_i \leq H_{jk} + B (1 - y_{ijk}) \quad \forall i \in N, \forall j \in M, \forall k \in P_j\\
		\end{split}\\
		\begin{split}
			\label{eq:16}
			-h_i \leq -H_{jk} + B (1 - y_{ijk}) \quad \forall i \in N, \forall j \in M, \forall k \in P_j\\
		\end{split}\\
		\begin{split}
			\label{eq:17}
			x_{ij} \leq \varphi_j \quad \forall i \in N, \forall j \in M\\
		\end{split}\\
		\begin{split}
			\label{eq:18}
			y_{ijk} \leq \gamma_{jk} \quad \forall i \in N, \forall j \in M, \forall k \in P_j\\
		\end{split}\\
		\begin{split}
			\label{eq:19}
			\gamma_{jk} \leq \sum_{i \in N} y_{ijk} \quad \forall j \in M, \forall k \in P_j\\
		\end{split}\\
		\begin{split}
			\label{eq:20}
			x_{ij} - x_{ij}^{'} \leq m_{ij} \quad \forall i \in N, \forall j \in M\\
		\end{split}\\
		\begin{split}
			\label{eq:21}
			x_{ij}^{'} - x_{ij} \leq m_{ij} \quad \forall i \in N, \forall j \in M\\
		\end{split}\\
		\begin{split}
			\label{eq:22}
			y_{ijk} - y_{ijk}^{'} \leq \omega_{ijk} \quad \forall i \in N, \forall j \in M, \forall k \in P_j\\
		\end{split}\\
		\begin{split}
			\label{eq:23}
			y_{ijk}^{'} - y_{ijk} \leq \omega_{ijk} \quad \forall i \in N, \forall j \in M, \forall k \in P_j\\
		\end{split}\\
		\begin{split}
			\label{eq:24}
			x_{ij}, y_{ijk}, \alpha_{ii'jk}, \varphi_j, \gamma_{jk}, m_{ij}, \omega_{ijk} \in \{0,1\}, z_{ijk} \in \mathbb{Z}^+, \beta_i \in \mathbb{Z}\\
		\end{split}
	\end{align}
	
	Eq. \eqref{eq:1} is designed to optimize the request acceptance by maximizing the sum of weights $a_i$ for each VM allocated. Eq. \eqref{eq:2} aims to consolidate resources by minimizing the number of active PMs and GPUs, using $b_j$ to weight each PM and $\gamma_{jk}$ to indicate active GPUs, thereby reducing active hardware and improving utilization. Eq. \eqref{eq:3} seeks to minimize both inter-machine and intra-machine migrations, thus enhancing system stability and efficiency. \( \delta_i \) is assigned a value of 0 for VMs that have just arrived and a value of 1 for VMs that are already residing in the system. Constraints set by Eq. \eqref{eq:4} and Eq. \eqref{eq:5} ensure that the CPU and RAM allocations for each PM do not exceed its respective capacities $C_j$ and $R_j$. Eq. \eqref{eq:6} guarantees that each VM is allocated to at most one PM, whereas Eq. \eqref{eq:7} ensures that the GI of each VM is assigned to at most one GPU. Eq. \eqref{eq:8} establishes that the GI of a VM can only be allocated if the VM itself is allocated, while Eq. \eqref{eq:9} ensures that $y_{ijk}$ is unset if $\text{VM}_i$ is not allocated to $\text{PM}_j$. Eq. \eqref{eq:10} and Eq. \eqref{eq:11} are designed to prevent overlapping GIs on the same GPU by ensuring proper ordering. $B$ is a large enough constant. Eq. \eqref{eq:12} to Eq. \eqref{eq:13} set bounds for the starting index $z_{ijk}$ of a GI on a GPU, where $g_i$ is the GI size of $\text{VM}_i$. The $\beta_i$ enables generation of GI size multiples. Eq. \eqref{eq:14} further restricts based on the maximum starting offset limit. Eq. \eqref{eq:15} to Eq. \eqref{eq:16} validate the compatibility between GI and GPU. Eq. \eqref{eq:17} to Eq. \eqref{eq:19} are used to determine $\varphi_j$ and $\gamma_{jk}$ ensuring that VMs are not assigned to powered-off PMs and GPUs. Eq. \eqref{eq:20} to Eq. \eqref{eq:23} are formulated to represent the live migration of VMs and the movement of their GIs. Finally, \ref{eq:24} defines the domains of all variables.
	
	\begin{table}[h]
		\centering
		\scriptsize
		\caption{Table of notations}
		\begin{tabular}{c|l}
			\hline
			\textbf{Notation} & \textbf{Description} \\
			\hline
			\( N \) & Set of virtual machines (VMs) \\
			\( M \) & Set of physical machines (PMs) \\
			VM\(_i\) & The \( i^{th} \) VM in the set \( N \) \\
			PM\(_j\) & The \( j^{th} \) PM in the set \( M \) \\
			\( P_j \) & Collection of GPUs in PM\(_j\) \\
			\( x_{ij} \) & Indicates allocation of VM\(_i\) to PM\(_j\) \\
			\( x_{ij}^{'} \) & Represents previous allocation state of VM\(_i\) to PM\(_j\) \\
			\( y_{ijk} \) & Represents assignment of GI of VM\(_i\) to the \( k^{th} \) GPU of PM\(_j\) \\
			\( y_{ijk}^{'} \) & Previous GI assignment of VM\(_i\) to GPU \( k \) of PM\(_j\) \\
			\( z_{ijk} \) & Starting offset of GI of VM\(_i\) in the \( k^{th} \) GPU of PM\(_j\) \\
			\( a_i \) & Weight associated with VM\(_i\) in the objective function \\
			\( b_j \) & Weight associated with PM\(_j\) in the objective function \\
			\( m_{ij} \) & Binary variable for changes in the allocation state of VM\(_i\) to PM\(_j\) \\
			\( \omega_{ijk} \) & Binary variable for changes in GI assignment of VM\(_i\) to GPU \( k \) of PM\(_j\) \\
			\( \delta_i \) & A parameter to omit newly arrived VM\(_i\) from migration minimization \\
			\( c_i \) & CPU requirement of VM\(_i\) \\
			\( C_j \) & CPU capacity of PM\(_j\) \\
			\( r_i \) & RAM requirement of VM\(_i\) \\
			\( R_j \) & RAM capacity of PM\(_j\) \\
			\( g_i \) & Size of the GI of VM\(_i\) \\
			\( \beta_i \) & Helps explore multiples of GI size as starting index for VM\(_i\) \\
			\( s_i \) & Last permissible index for the GI of VM\(_i\) \\
			\( B \) & A large enough constant \\
			\( \varphi_j \) & Binary variable indicating if PM\(_j\) is powered-on \\
			\( H_{jk} \) & Characteristic of the \( k^{th} \) GPU in PM\(_j\) \\
			\( h_i \) & Characteristic of the GI of VM\(_i\) \\
			\( \alpha_{ii'jk} \) & Binary variable related to the ordering of GIs on GPUs \\
			\hline
		\end{tabular}
		\label{tab:notations}
	\end{table}

\begin{table}[ht]
	\centering
	\caption{The values of $g_i$, $s_i$, and $h_i$ for NVIDIA A100 MIG profiles.}
	\begin{tabular}{llll}
		\hline
		Profile & $g_i$ & $s_i$ & $h_i$ \\ \hline
		1g.5gb  & 1     & 6     & 100   \\
		1g.10gb & 2     & 6     & 100   \\
		2g.10gb & 2     & 4     & 100   \\
		3g.20gb & 4     & 4     & 100   \\
		4g.20gb & 4     & 0     & 100   \\
		7g.40gb & 8     & 0     & 100   \\ \hline
	\end{tabular}
	\label{tab:model-params}
\end{table}

	Table \ref{tab:model-params} presents example parameters for various NVIDIA A100 MIG profiles. In this illustration, it is assumed that \( H_{jk} = 100 \) for every \( j \in M \) and \( k \in P_j \), when \( k \) corresponds to an A100 GPU. The profiles with identical memory requirements are given similar \( g_i \) values, although their compute engine needs may vary. These example settings of \( g_i \) and \( s_i \) demonstrate how the allocation can be managed.
	
	The weights \( a_i \) are important for aligning VM allocation with provider policies and mitigating the tendency to accept smaller VMs preferentially. They allow the provider to rank VMs based on predefined criteria. In contrast, the weights \( b_j \) enable the provider to modify the selection priority of physical machines. This modification can be aimed at objectives like lowering the selection of high power-consuming machines. For example, suppose a provider decides to prioritize larger VMs due to their higher revenue potential and to deprioritize power-hungry machines. In this scenario, larger VMs might be assigned higher \( a_i \) values (e.g., \( a_i = 5 \) for large VMs and \( a_i = 1 \) for small VMs), and machines with higher power consumption might be assigned higher \( b_j \) values (e.g., \( b_j = 5 \) for high power-consuming machines and \( b_j = 1 \) for energy-efficient machines). This way, the optimization algorithm would prefer allocating resources to larger VMs and using more energy-efficient machines.
	
	The parameter \( \delta_i \), initialized as zero for new VMs, prevents migration for these VMs. Conversely, a higher value inhibits the migration of VMs of interest. For instance, an elevated value assigned to a previously migrated VM reduces the likelihood of frequent VM preemptions. Moreover, inter-machine migrations, denoted by \( m_{ij} = 1 \), trigger intra-machine GPU reassignments (\( \omega_{ijk} \)), but intra-machine adjustments do not necessarily result in machine switches. This model characteristic prioritizes less disruptive and cost-effective intra-machine modifications over inter-machine migrations. This preference aligns with other objectives such as efficient VM allocation and minimizing active hardware usage, contributing to overall resource optimization and reducing fragmentation.

	\section{The Proposed Method}
	\label{s:methodology}
	
	The problem in Section \ref{s:formulation} combines elements of the Knapsack and Bin Packing problems, with added complexities. Eq. \eqref{eq:1} aims to maximize request acceptance within limited machine capacity, like the Knapsack problem, while Eq. \eqref{eq:2} minimizes hardware usage for requests, similar to the Bin Packing problem. This problem is complicated by its online stochastic nature, requiring decision-making without full knowledge of future demands. It differs from the static Knapsack and Bin Packing problems by needing dynamic management and potential relocation of VMs as in Eq. \eqref{eq:3}, adding temporal complexity. Given the NP-hardness of the Knapsack and Bin Packing problems and the complexity of solving ILP formulations, a direct solution is impractical. Instead, the problem is broken down into multiple steps, each solved individually, to offer more feasible solutions within its complexity constraints.
	
	\begin{figure}[h]
		\centering
		\includegraphics[trim=0.7cm 0.4cm 0.9cm 0.4cm, clip,, width=0.8\columnwidth]{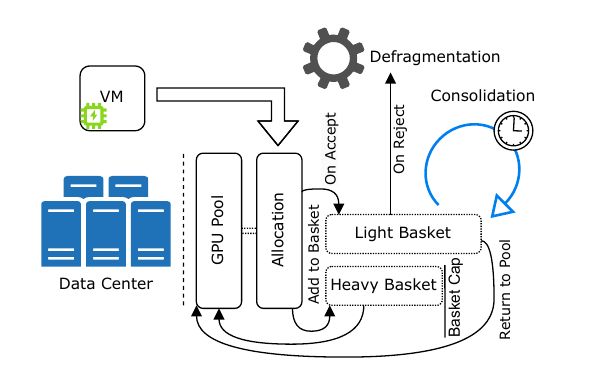}
		\caption{Components of the multi-stage GRMU placement}
		\label{fig:grmu}
	\end{figure}
	
	\subsection{GPU Resource Management Unit (GRMU)}
	
	To address the problem, we present GRMU, depicted in Fig.~\ref{fig:grmu}, which facilitates efficient workload-to-resource mapping aligned with specific objective functions. This section explores the GRMU algorithms for GPU resource allocation and optimization, highlighting only GPUs and key code sections for clarity. Algorithm \ref{alg:init} outlines the GRMU initialization process, where GPUs from data center hosts are organized into a resource pool. The algorithm sets a capacity limit for the heavy basket and assigns a unique global index to each GPU to enable an orderly first-fit search. The pooled GPUs are then divided into heavy and light groups, each starting with one empty GPU, a method we refer to \emph{Dual-Basket Pooling} in this paper.

\begin{algorithm}[h]
	\caption{\small GRMU Initialization\protect\footnotemark}
	\footnotesize
	\label{alg:init}
	\begin{algorithmic}[1]
		\STATE \texttt{pool, heavyBasket, lightBasket} $\leftarrow \emptyset$
		\STATE \texttt{heavyBasketCapacity} $\leftarrow$ \textsc{<predefinedValue>}
		\STATE \texttt{globalIndex} $\leftarrow 0$
		
		\STATE \textbf{for} \texttt{host} $\in$ \texttt{hosts} \textbf{do}
		\STATE \quad \textbf{for} \texttt{gpu} $\in$ \texttt{host.gpus} \textbf{do}
		\STATE \quad\quad \texttt{gpu.globalIndex} $\leftarrow$ \texttt{globalIndex}
		\STATE \quad\quad \texttt{globalIndex} $\leftarrow$ \texttt{globalIndex} $+ 1$
		\STATE \quad\quad \textsc{Add}\texttt{(pool, gpu)}
		
		\vspace{0.1cm}
		
		\STATE \textsc{Add}\texttt{(heavyBasket,} \textsc{Get}\texttt{(pool))}
		\STATE \textsc{Add}\texttt{(lightBasket,} \textsc{Get}\texttt{(pool))}
	\end{algorithmic}
\end{algorithm}

		Algorithm \ref{alg:allocation} demonstrates the allocation policy for handling newly arrived VMs. For each VM, the appropriate GPU basket is determined: the heavy basket is chosen for VMs with a 7g.40gb MIG profile, while the light basket is used for other profiles. The algorithm then scans the GPUs in the selected basket to attempt allocation (see the \textsc{Assign} function in Algorithm \ref{alg:cc}). GRMU limits heavy basket sizes to prevent 7g.40gb profiles from monopolizing entire GPUs, ensuring availability for the light basket. If no GPU in the selected basket can host the VM profile and the basket size allows, a new GPU is taken from the pool, added to the basket, and allocated to the VM.

	\begin{algorithm}[h]
		\caption{\small VM Allocation}
		\footnotesize
		\label{alg:allocation}
		\begin{algorithmic}[1]
			\STATE \textbf{for} \texttt{vm} $\in$ \texttt{vms} \textbf{do}
			\STATE \quad \textbf{if} \texttt{vm.gpu} $\neq$ \texttt{7g.40gb} \textbf{then}
			\STATE \quad\quad \texttt{basket} $\leftarrow$ \texttt{lightBasket}
			\STATE \quad\quad \texttt{basketCapacity} $\leftarrow$ \texttt{numGpus} $-$ \texttt{heavyBasketCapacity}
			\STATE \quad \textbf{else}
			\STATE \quad\quad \texttt{basket} $\leftarrow$ \texttt{heavyBasket}
			\STATE \quad\quad \texttt{basketCapacity} $\leftarrow$ \texttt{heavyBasketCapacity}
			
			\STATE \quad \texttt{success} $\leftarrow$ \textbf{False}
			
			\STATE \quad \textbf{for} \texttt{gpu} $\in$ \texttt{basket} \textbf{do}
			\STATE \quad\quad \textbf{if} \textsc{Assign}\texttt{(vm.profile, gpu)} \textbf{then}
			\STATE \quad\quad\quad \texttt{success} $\leftarrow$ \textbf{True}
			\STATE \quad\quad\quad \textbf{break}
			
			\STATE \quad \textbf{if not} \texttt{success} \textbf{and} $|\texttt{basket}| \leq \texttt{basketCapacity}$ \textbf{then}
			\STATE \quad\quad \texttt{gpu} $\leftarrow$ \textsc{Get}\texttt{(pool)}
			\STATE \quad\quad \textbf{if} \texttt{gpu} $\neq \emptyset$ \textbf{then}
			\STATE \quad\quad\quad \textsc{Add}\texttt{(basket, gpu)}
			\STATE \quad\quad\quad \textsc{Assign}\texttt{(vm.profile, gpu)}
		\end{algorithmic}
	\end{algorithm}

	When GRMU notices the allocation results and finds that any VMs have been rejected, it triggers a defragmentation process within the light basket to free up capacity. While GPUs in the heavy basket remain fully utilized, the light basket may have fragmented resources. As shown in Algorithm \ref{alg:frag_defrag_functions}, the defragmentation process calculates a fragmentation value for each GPU in the light basket to identify the most fragmented GPU for reorganization. The GPU blocks are then rearranged to enhance the GPU $CC$ metric through a process called \emph{Intra-GPU Migration}. 
	
	When MIG profiles are initially placed on a GPU, they are positioned on blocks that maximize the GPU $CC$ value. As explained in Section \ref{ss:analysis}, a 1g.5gb profile is placed on block 6. The second 1g.5gb profile is positioned on block 4. However, if the first profile leaves before the second, the GPU $CC$ value would remain in a suboptimal state unless the remaining profile is moved to block 6, where the configuration achieves the maximum $CC$. As shown in Algorithm \ref{alg:frag_defrag_functions}, once the most fragmented GPU is identified, an empty mock GPU is created to determine which VMs require intra-GPU migration. Then, VMs from the fragmented GPU are allocated onto the mock GPU using the default MIG profile placement. This is not an actual allocation but serves to determine which VMs would be assigned to different blocks compared to their current assignments. The \textsc{Relocated} function then compares the placement of VMs in the original fragmented GPU and the mock GPU to identify any VM relocations needed to increase the $CC$ value.
	
\begin{algorithm}[h]
	\caption{\small Defragmentation and Fragmentation}
	\footnotesize
	\label{alg:frag_defrag_functions}
	\begin{algorithmic}[1]
		
		\STATE \textbf{Function} \textsc{Defragmentation}(\texttt{lightBasket})
		\STATE \quad\texttt{gpu} $\leftarrow$ \textsc{Max}\texttt{(lightBasket}, \textsc{Fragmentation)}
		\STATE \quad \texttt{mockGpu} $\leftarrow$ \{0, 1, 2, 3, 4, 5, 6, 7\}
		
		\STATE \quad \textbf{for} \texttt{vm} $\in$ \texttt{gpu.vms} \textbf{do}
		\STATE \quad \quad \textsc{Assign}\texttt{(vm.profile, mockGpu)}
		
		\STATE \quad \texttt{vmsToMigrate} $\leftarrow$ \textsc{Relocated}\texttt{(gpu, mockGpu)}
		\STATE \quad \textsc{IntraMigrate}\texttt{(vmsToMigrate, gpu)}
		
		\vspace{0.1cm}
		
		\STATE \textbf{Function} \textsc{Fragmentation}(\texttt{gpu})
		\STATE \quad \texttt{fragVal} $\leftarrow 0$
		\STATE \quad \texttt{gpu$^\prime$} $\leftarrow$ \textsc{Copy}\texttt{(gpu)}
		\STATE \quad \textbf{for} \texttt{profile} $\in \{ p \in \textsc{Profiles} \mid \textsc{Size}\texttt{(p)} \leq |\texttt{gpu$^\prime$}| \}$ \textbf{do}
		\STATE \quad \quad \quad \textbf{for} \texttt{start} $\in$ \texttt{startBlocks[profile]} \textbf{do}
		\STATE \quad\quad\quad\quad \texttt{blocks} $\leftarrow \{ \texttt{start} + i \mid i < \textsc{Size}\texttt{(profile)} \}$
		\STATE \quad\quad\quad\quad \textbf{if} \texttt{blocks} $\subseteq$ \texttt{gpu$^\prime$} \textbf{then}
		\STATE \quad\quad\quad\quad\quad \texttt{gpu$^\prime$} $\leftarrow$ \texttt{gpu$^\prime$} $\setminus$ \texttt{blocks}
		
		\STATE \quad\quad\quad \texttt{fragVal} $\leftarrow$ \texttt{fragVal} $+ \left( |\texttt{gpu$^\prime$}| / \textsc{Size}\texttt{(profile)} \right)$
		
		\STATE \quad\textbf{return} \texttt{fragVal}
		
	\end{algorithmic}
\end{algorithm}
	
	To determine the most fragmented GPU, the \textsc{Fragmentation} function assigns a fragmentation value to each GPU. It calculates this value by generating a score based on the distribution of occupied and free blocks. For each MIG profile that is smaller than or equal to the available GPU free blocks, the algorithm attempts to remove as much of the profile as possible and adds the ratio of remaining free blocks to profile size to the fragmentation value. The final score is the sum of these values for all relevant profiles, representing the amount of unusable space in the GPU that could not be allocated to profiles.
	
	\footnotetext{The symbol `\$' denotes each element in the list.}
	
	\begin{algorithm}[h]
		\caption{\small Light Basket Consolidation}
		\footnotesize
		\label{alg:consolidation}
		\begin{algorithmic}[1]
			\STATE \texttt{gpus} $\leftarrow \{ \texttt{gpu} \in \texttt{lightBasket} \mid \textsc{HalfFull}\texttt{(gpu)} \land \textsc{SingleProfile}\texttt{(gpu)} \}$

			\STATE \textbf{for} \texttt{sourceGpu} $\in$ \texttt{gpus} \textbf{do}
			\STATE \quad \texttt{targetGpu} $\gets$ \textbf{any} \ \texttt{gpus} $\setminus$ \{\texttt{sourceGpu}\}
			
			\STATE \quad \textsc{InterMigrate}\texttt{(sourceGpu} \textbf{to} \texttt{targetGpu)}
			\STATE \quad \texttt{gpus} $\leftarrow$ \texttt{gpus} $\setminus \{\texttt{sourceGpu}\}$
			\STATE \quad \textsc{Remove}\texttt{(lightBasket, sourceGpu)}
			\STATE \quad \textsc{Add}\texttt{(pool, sourceGpu)}
			
		\end{algorithmic}
	\end{algorithm}

		To further optimize resource allocation, the system conducts regular \emph{Inter-GPU Migration} on light basket GPUs, targeting half-empty GPUs with a single 3g.20gb or 4g.20gb profiles. Algorithm \ref{alg:consolidation} identifies these GPUs and consolidates workloads so that one GPU becomes fully utilized while the other is emptied. Freed GPUs are removed and returned to the pool by \texttt{globalIndex} order.

		To address conflicting objective functions, the GRMU employs several strategies. \emph{Dual-Basket Pooling} with per-basket GPU limits improves the acceptance rate of smaller profiles; however, the cap must be carefully configured to avoid idle resources if no small GPU requests arrive. The \emph{First-Fit} strategy promotes consolidation by efficiently assigning workloads to GPUs. Through \emph{Intra-GPU Migration}, GPUs can free up capacity for future requests. Moreover, \emph{Inter-GPU Migration} consolidates workloads onto fully utilized GPUs, allowing freed GPUs to rejoin the pool for potential future allocations. Notably, consolidation contributes to improved resource utilization. By applying intra-GPU migration across all profiles of only the most fragmented GPU and limiting inter-GPU consolidation to large profiles, the system minimizes migrations while maintaining effectiveness. The interval for initiating consolidation must be carefully set.
	
	\section{Evaluation}
	\label{s:evaluation}
	
	The evaluation of the effectiveness of methods in online algorithms is commonly performed through a measure known as competitiveness, where the output of an offline variant of the problem serves as a benchmark for comparison.Nonetheless, as noted in Section \ref{s:methodology}, the substantial complexity and scale of the problem make this approach unsuitable for our case, as even a solver cannot handle it within a viable timeframe, even in limited-scale scenarios.
	
	We use Cloudy \cite{cloudy} to conduct our simulations. The placement process operates on two hierarchical levels. At the upper level, VM placement determines the sequence in which hosts and their respective GPUs are traversed. At the lower level, MIG profile placement assigns specific GPU blocks to the corresponding VM. The policies in this section, including GRMU, operate at the first level, while the second level relies on the default MIG profile placement by NVIDIA, which is fixed and cannot be overridden. However, understanding its behavior allows the upper level MIG-aware VM scheduler to make more informed decisions.
	
	\subsection{Workload}
	\label{ss:workload}
	
	To construct the workload for our simulations, we use the Alibaba GPU cluster trace from 2023 \cite{noauthor_clusterdatacluster-trace-gpu-v2023_nodate}, treating nodes as hosts and pods as VMs. We remove arrival time outliers using the interquartile range (IQR) method \cite{SMITI2020100306}, where Q1 and Q3 are the 25th and 75th percentiles, and outliers are values outside 1.5 times the IQR from Q1 and Q3. Each node, equipped with one to eight GPUs, will have each GPU mapped to an NVIDIA A100, and each pod's GPU requirement will be mapped to a single MIG profile.
	
	To map a pod GPU to a MIG profile, we first calculate the pod's total GPU requirement by multiplying the number of GPUs needed by the fraction of each GPU required. Pods requiring more than one complete GPU, which are less than 1\%, are excluded as they are unsupported by our simulator. We normalize the total GPU requirement \( u \) by dividing it by the maximum GPU requirement across all pods, as shown in Eq. \eqref{eq:normalized_gpu_requirement}. Each MIG profile is characterized by its compute and memory units, combined into a value \( U_k \) as in Eq. \eqref{eq:combined_value}. These profile values are normalized by the maximum combined value, as per Eq. \eqref{eq:normalized_profile_value}. For each pod, we assign the MIG profile whose normalized value is closest to the pod's normalized GPU requirement by solving Eq. \eqref{eq:assigned_profile}.
	
	\begin{align}
		\hat{u} &= \frac{u}{\max(u)} \label{eq:normalized_gpu_requirement} \\
		U_k &= \textit{compute}_k \times \textit{memory}_k \label{eq:combined_value} \\
		\hat{U}_k &= \frac{U_k}{\max(U_k)} \label{eq:normalized_profile_value} \\
		k^* &= \arg\min_k \left| \hat{U}_k - \hat{u} \right| \label{eq:assigned_profile}
	\end{align}
	
\begin{figure}[h]
		\centering
		\includegraphics[width=0.8\columnwidth]{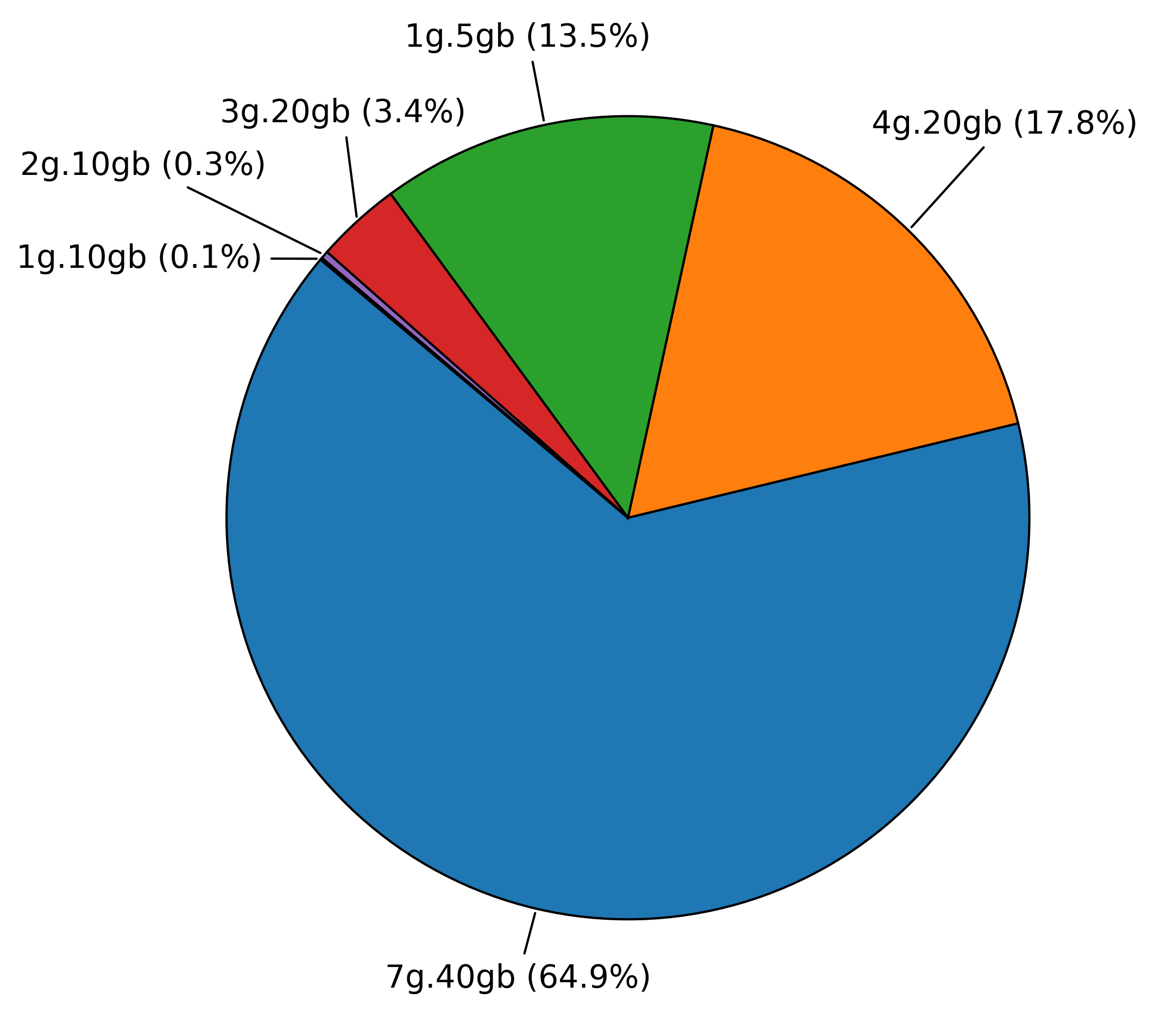}
		\caption{Distribution of profiles in the workload}
		\label{fig:workload-histogram}
	\end{figure}

	Here, \( k^* \) represents the index of the profile assigned to the pod, determined by finding the profile \( k \) whose normalized combined value \( \hat{C}_k \) is closest to the pod's normalized GPU requirement \( \hat{g} \). This method ensures that the GPU requirements of pods are matched to the closest available profiles based on normalized values. Finally, there will be a total of $1,213$ GPU-equipped hosts and $8,063$ MIG-enabled VMs. Fig.~\ref{fig:workload-histogram} illustrates the distribution of MIG profiles within the VMs.
	
	\subsection{Stepwise Analysis}
	
	This subsection evaluates the individual components of GRMU and their isolated impacts on key metrics, while also selecting good enough values for key parameters like heavy basket capacity and consolidation interval. The parameters are tuned per workload and must be adjusted for each provider pattern. We initially disable consolidation and defragmentation, enabling them sequentially to evaluate their effects.
	
	\begin{figure}[h]
		\centering
		\includegraphics[width=1\columnwidth]{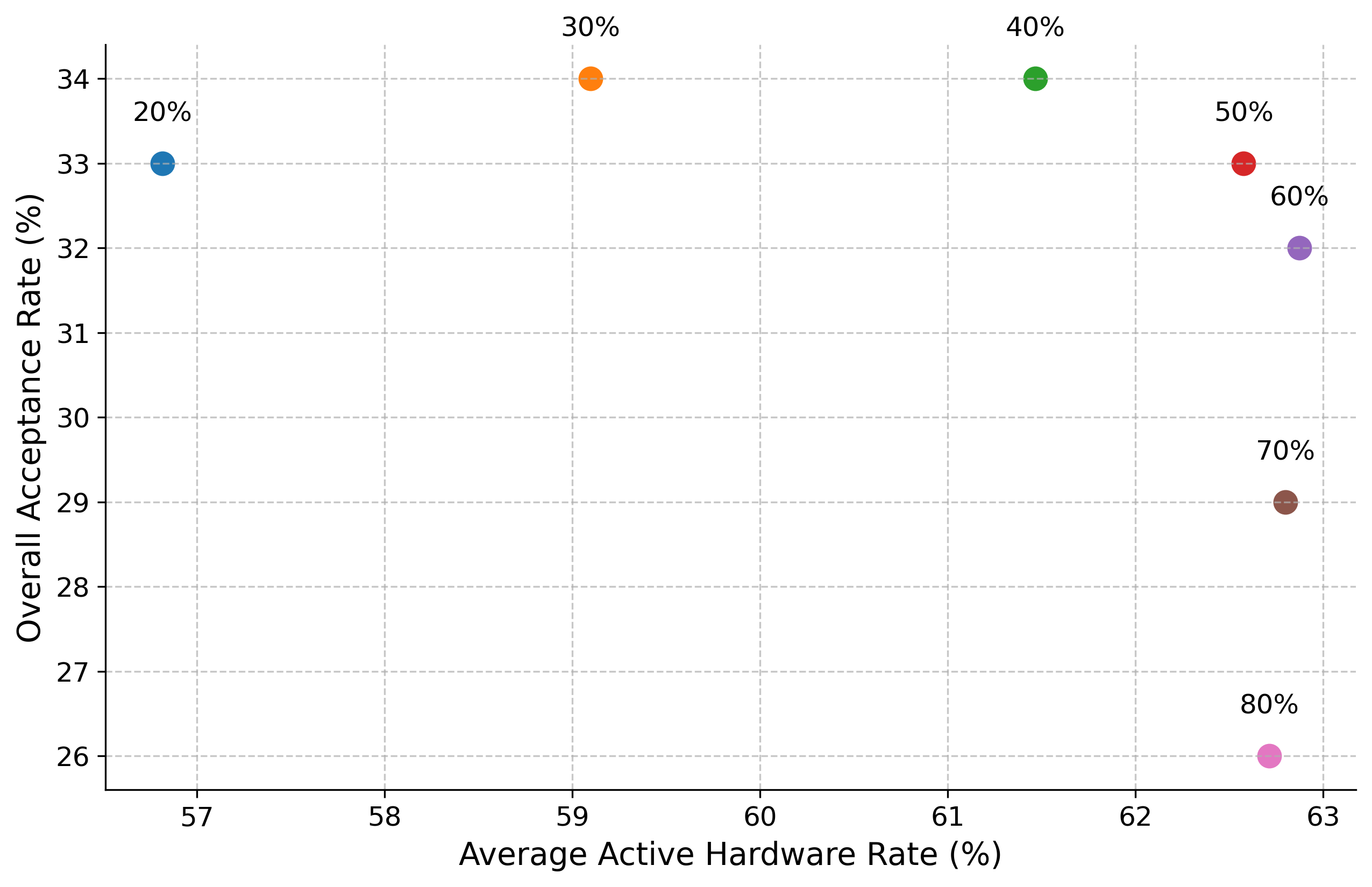}
		\caption{Impact of heavy basket capacity with annotations showing the GPUs reserved for it}
		\label{fig:dualbasket}
	\end{figure}
	
	\subsubsection{Heavy Basket Capacity}
	
	To find the optimal heavy basket capacity, Fig.~\ref{fig:dualbasket} shows how the Average Active Hardware Rate and the Overall Acceptance Rate across basket capacities ranging from 20\% to 80\%. The Average Active Hardware Rate is the mean of hourly active hardware rates, while the Overall Acceptance Rate is the final acceptance rate at the end of the simulation. Here, a 20\% capacity means only 20\% of all GPUs are reserved for the heavy basket. The defragmentation and consolidation components of GRMU are disabled to isolate the impact of heavy basket capacity; as there are no migrations, they are omitted from the diagram. Higher overall acceptance rates indicate better request handling, while lower average active hardware rates suggest more efficient resource use. The plot shows that increasing basket capacity generally raises active hardware usage but improves acceptance rate up to a certain point. The 30\% basket capacity achieves a good balance, with a high acceptance rate and relatively low active hardware usage, making it a strong choice for maximizing request acceptance and efficient resource utilization.
	
	\begin{figure}[h]
		\centering
		\includegraphics[width=1\columnwidth]{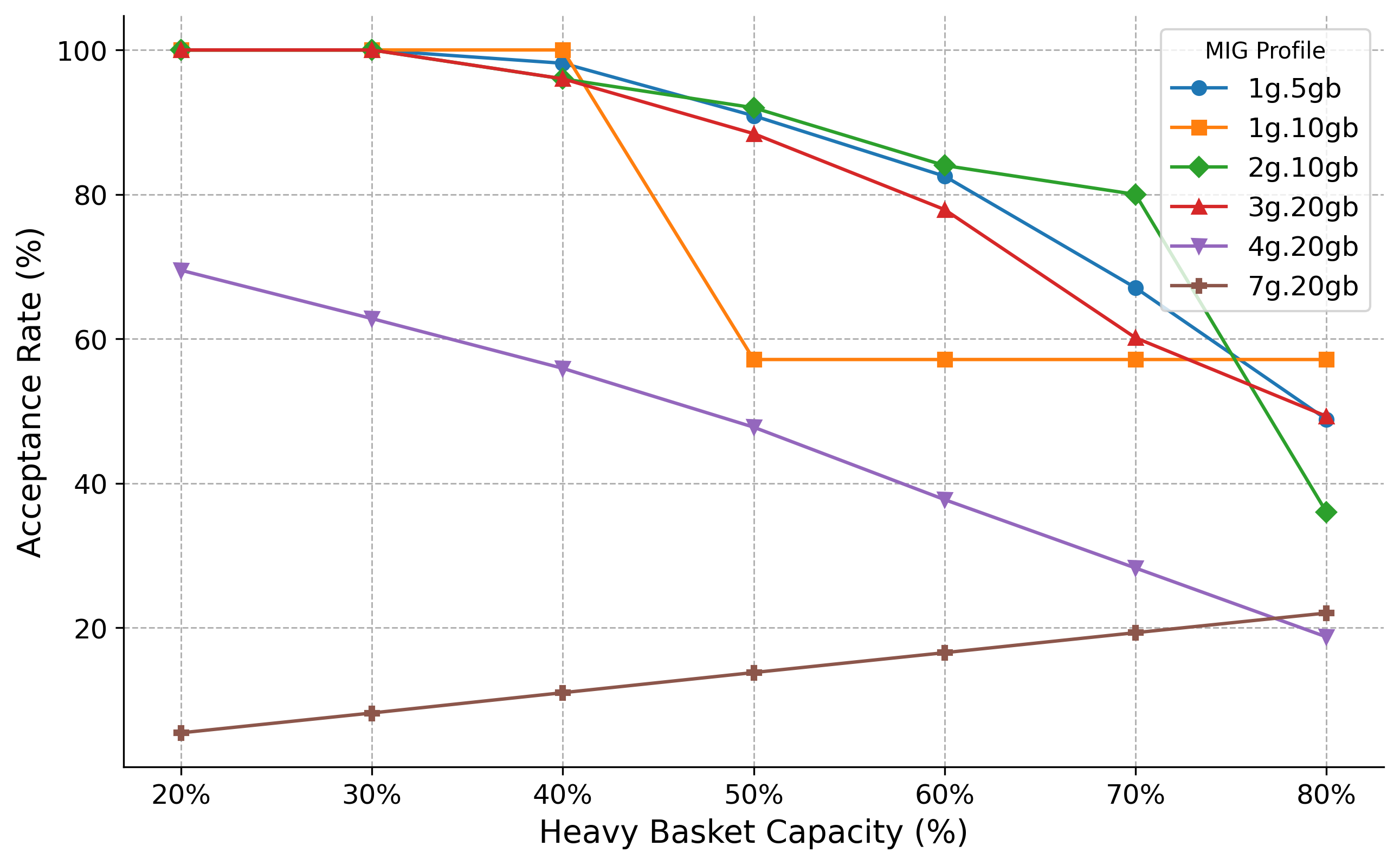}
		\caption{Acceptance of requested profiles across varying heavy basket capacities}
		\label{fig:dualbasket-workload}
	\end{figure}
	
	\begin{figure}[h]
		\centering
		\includegraphics[width=1\columnwidth]{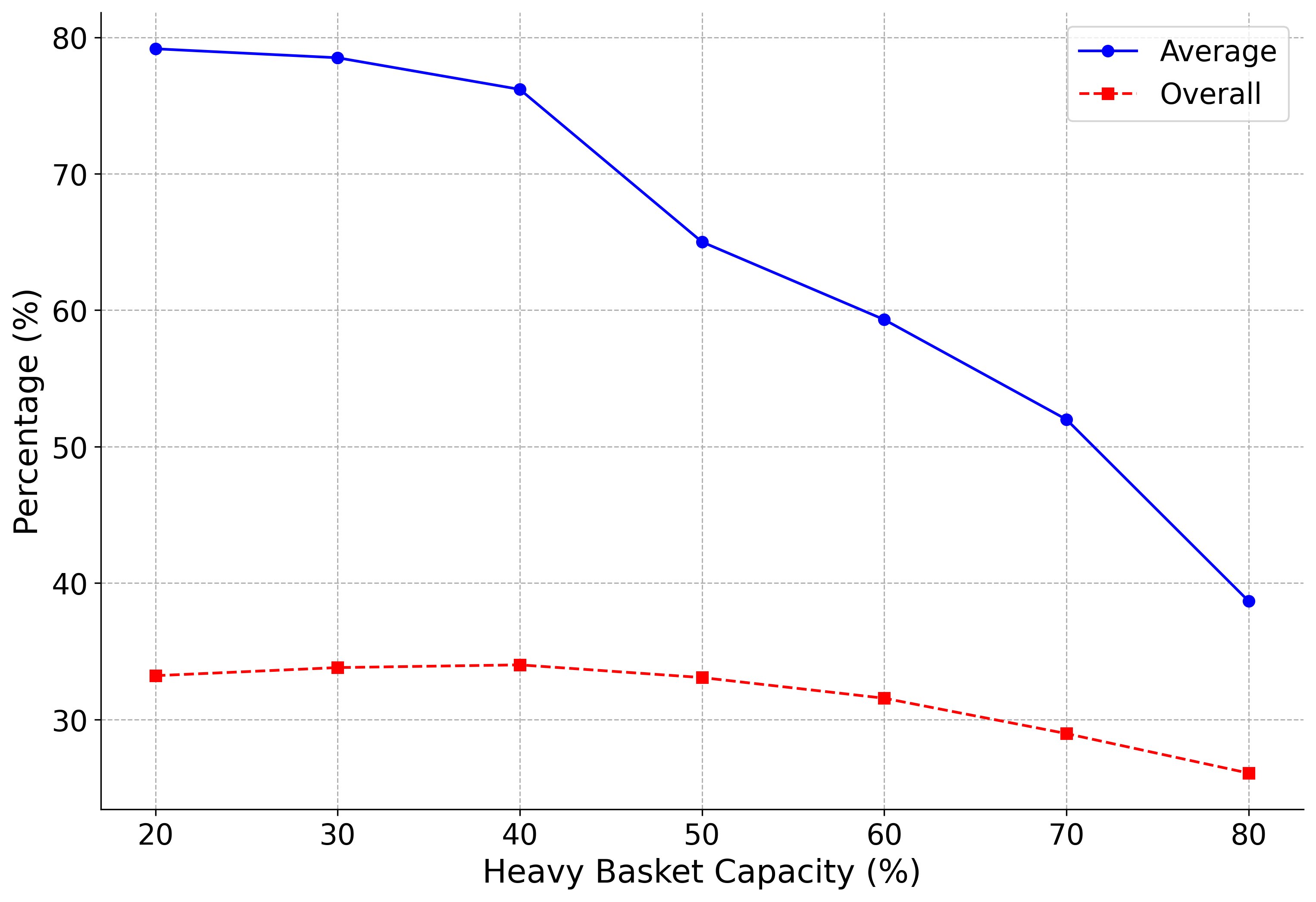}
		\caption{Comparison of average and overall acceptance rates across varying heavy basket capacities}
		\label{fig:dualbasket-workload-stats}
	\end{figure}
	
	Fig.~\ref{fig:dualbasket-workload} illustrates the acceptance rate of each MIG profile across different heavy basket capacities. Each point represents the proportion of accepted requests relative to the total requests for that profile at a given capacity. As the heavy basket capacity increases, acceptance for the 7g.40gb profile rises, while others decline due to limited allocation space. As expected, a decrease in the number of small profiles leads to a lower overall acceptance rate, as shown by the red line in Fig.~\ref{fig:dualbasket-workload-stats}. The blue line in Fig.~\ref{fig:dualbasket-workload-stats} represents the average acceptance rate across profiles for each basket capacity, calculated as the mean of the corresponding points in Fig.~\ref{fig:dualbasket-workload}. A basket capacity of 30\% appears to provide a favorable balance between overall and average acceptance rates. This analysis assumes that all profiles receive equal priority from the provider.
	
	\subsubsection{Consolidation Interval}

\begin{figure}[h]
		\centering
		\includegraphics[width=1\columnwidth]{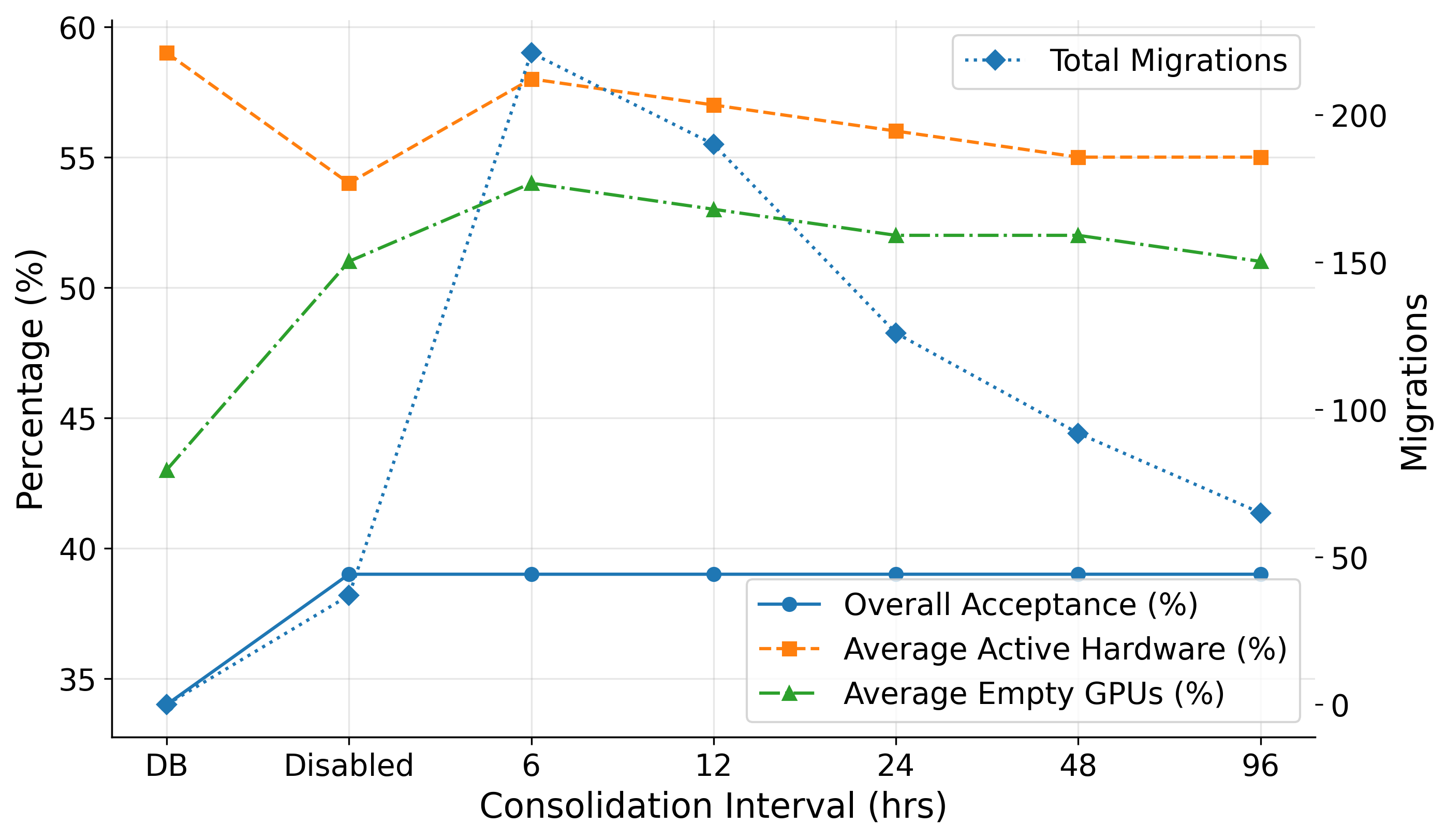}
		\caption{Objective function values for different consolidation intervals}
		\label{fig:consolidation-intervals}
	\end{figure}

	The next step is to tune the consolidation interval. Simulations are conducted with all GRMU components enabled across different consolidation intervals. Fig. \ref{fig:consolidation-intervals} presents the results. DB denotes the scenario with only \emph{DUAL-Basket Pooling} (heavy basket capacity at 30\%), while \emph{Disabled} corresponds to active defragmentation without consolidation. The remaining points represent the various consolidation intervals of 6, 12, 24, 48, and 96 hours. 
	
	To find the optimal interval, we aim to maximize acceptance while minimizing active resources and migrations (see Eq. \eqref{eq:1} to Eq. \eqref{eq:3}). This value may vary based on provider priorities and workload patterns. Interestingly, for the given workload, the \emph{Disabled} point—where consolidation is off and only defragmentation is active—exhibits these desired characteristics. While this point maintains fewer empty GPUs on average, the arrival pattern of requests prevents their utilization. Shorter intervals lead to more migrations, which consolidate free GPU blocks and result in more empty GPUs. This demonstrates the advantage of the consolidation component and highlights the importance of tuning it based on workload. For the selected workload, we proceed with \emph{Disabled} consolidation. If a provider prioritizes energy over performance, other points may be considered.
	
	\subsection{Comparison}
	
	We compare the GRMU policy with four VM placement policies: First-Fit (FF), Max Configuration Capacity (MCC), Max Extended Configuration Capacity (MECC), and Best-Fit (BF). Each of these policies offers distinct advantages, with varying approaches to resource allocation and optimization, depending on the specific needs of the system. Each policy is outlined below:
	
	\begin{enumerate}
		\item \textbf{FF Policy}:  
		It sequentially scans hosts and their GPUs to find a suitable match for a VM allocation request. Upon finding a compatible resource, the request is placed immediately. This policy is widely adopted due to its simplicity and effectiveness.
		
		\item \textbf{MCC Policy}:  
		It evaluates all available GPUs in the data center and selects the one that, after allocation, results in the highest CC value. The procedure is outlined in Algorithm \ref{alg:max_cc_allocation}.
		
		\begin{algorithm}[h]
			\caption{\small Max CC VM Allocation}
			\footnotesize
			\label{alg:max_cc_allocation}
			\begin{algorithmic}[1]
				\STATE \textbf{for} \texttt{vm} $\in$ \texttt{vms} \textbf{do}
				\STATE \quad \texttt{bestGpu} $\leftarrow \emptyset$
				\STATE \quad \texttt{maxCC} $\leftarrow -1$
				
				\STATE \quad \textbf{for} \texttt{gpu} $\in$ \texttt{pool} \textbf{do}
				\STATE \quad\quad \textbf{if} \textsc{Assign}\texttt{(vm.profile, gpu)} \textbf{then}
				\STATE \quad\quad\quad\quad \texttt{currentCC} $\leftarrow \textsc{GetCC}(\texttt{gpu})$
				\STATE \quad\quad\quad\quad \textbf{if} \texttt{currentCC} $>$ \texttt{maxCC} \textbf{then}
				\STATE \quad\quad\quad\quad\quad \texttt{bestGpu} $\leftarrow \texttt{gpu}$
				\STATE \quad\quad\quad\quad\quad \texttt{maxCC} $\leftarrow \texttt{currentCC}$
				\STATE \quad\quad\quad \textsc{UnAssign}\texttt{(vm.profile, gpu)} \texttt{// Undo Assign}
				\STATE \quad \textbf{if} \texttt{bestGpu} $\neq \textsc{None}$ \textbf{then}
				\STATE \quad\quad \textsc{Assign}\texttt{(vm.profile, bestGpu)}
			\end{algorithmic}
		\end{algorithm}
		
		\item \textbf{MECC Policy}:  
		It is a variation of the MCC algorithm, where \textsc{GetCC} in line 6 is replaced with \textsc{GetECC}. Algorithm \ref{alg:get_ecc} provides the details of the \textsc{GetECC} function. This approach calculates the CC for a GPU by multiplying the probability of a profile's appearance by its count during summation. The probabilities are derived from a $n$-hour time window, ensuring a dynamically responsive resource allocation strategy.
		
		\begin{algorithm}[h]
			\caption{\small Expected Configuration Capability}
			\footnotesize
			\label{alg:get_ecc}
					\begin{algorithmic}[1]
		\STATE \textbf{Function} \textsc{GetECC}\texttt{(G)}
		\STATE \quad \texttt{ECC} $\leftarrow 0$
		\STATE \quad \textbf{for} \texttt{profile, starts} $\in$ \texttt{startBlocks} \textbf{do}
		\STATE \quad\quad \texttt{CC} $\leftarrow 0$
		\STATE \quad\quad \textbf{for} \texttt{start} $\in$ \texttt{starts} \textbf{do}
		\STATE \quad\quad\quad \texttt{blocks} $\leftarrow \{ \texttt{start} + i \mid i < \textsc{Size(profile)} \}$
		\STATE \quad\quad\quad \textbf{if} \texttt{blocks} $\subseteq$ \texttt{G} \textbf{then}
		\STATE \quad\quad\quad\quad \texttt{CC} $\leftarrow$ \texttt{CC} + 1
		\STATE \quad\quad \texttt{ECC} $\leftarrow$ $\texttt{P(profile)}\times\texttt{CC}$
		\STATE \quad \textbf{return} \texttt{ECC}
			\end{algorithmic}
		\end{algorithm}

		We tested \( n \) values of 1, 12, 24, 48, and 96 hours to predict the most probable profile in the current workload, calculating prediction errors for each. The \( n = 24 \) hours window resulted in the lowest error rate of 35\%. Thus, 24 hours was picked as the look-back period for profile predictions in the MECC policy.
		
		\item \textbf{BF Policy}:  
		The BF policy identifies all GPUs in the data center that are capable of accommodating the given VM request. It then selects the GPU that minimizes the remaining unallocated blocks, aiming for the most efficient use of GPU resources.
		
		\item \textbf{GRMU Policy}:
		It is described in Section \ref{s:methodology}, and for the current workload, as derived in Section \ref{ss:analysis}, the heavy basket capacity is set to 30\% of the GPU pool, while the consolidation interval is disabled.
		
		\end{enumerate}
		
		\subsubsection{Acceptance Rates}
		
\begin{figure}[h]
			\centering
			\includegraphics[width=1\columnwidth]{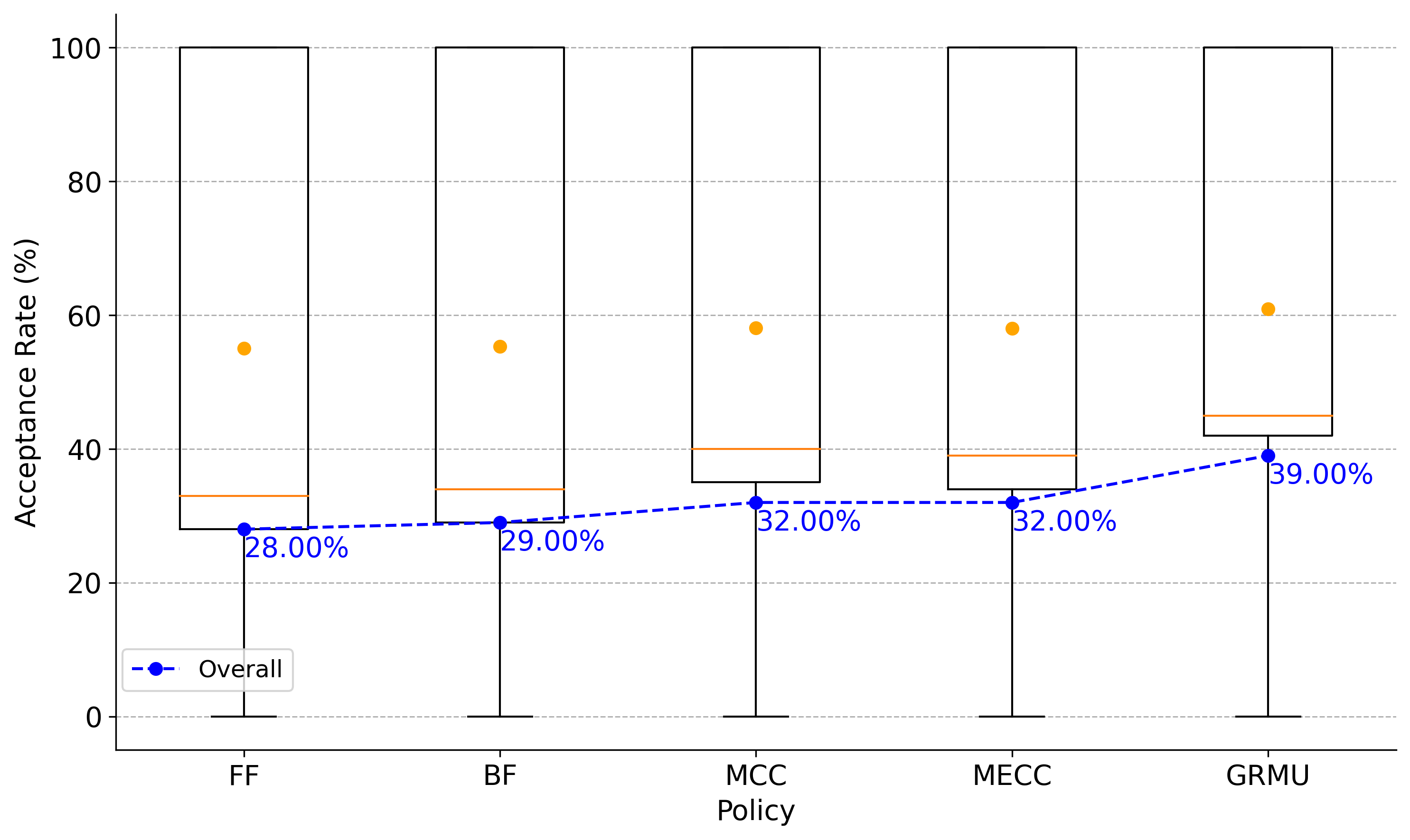}
			\caption{Acceptance rates by policy}
			\label{fig:acceptance-rates}
		\end{figure}

\begin{figure}[h]
			\centering
			\includegraphics[width=1\columnwidth]{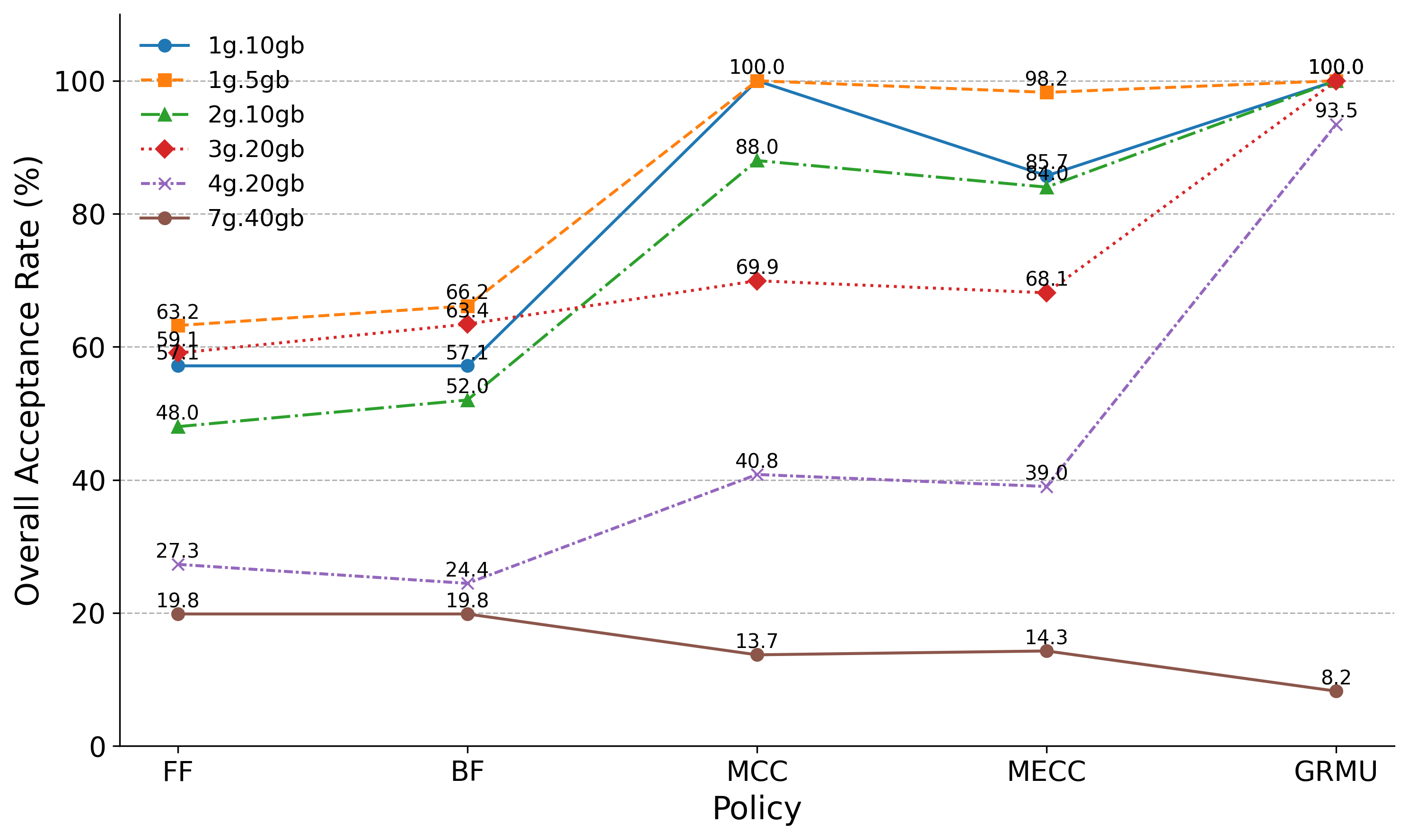}
			\caption{Acceptance rates per policy across GPU profiles}
			\label{fig:profiles-acceptance-rates}
		\end{figure}

		Fig. \ref{fig:acceptance-rates} presents the hourly acceptance rates throughout the simulation, including the overall rate at its conclusion. Fig. \ref{fig:profiles-acceptance-rates} provides a breakdown of acceptance rates for each policy by profile. Notably, GRMU outperforms all other policies across all profiles, except for the 7g.40gb profile, where the basket size was restricted. 
		
		The MECC performs similarly to the MCC policy. However, since its effectiveness relies on workload prediction, it only outperforms MCC in the 7g.40gb profile. In this case, MECC makes a better prediction due to the abundance of the profile (see Fig. \ref{fig:workload-histogram}), leading to slightly better performance.
		
		Compared to the second-best policy, MCC, GRMU shows improvements of 1.14×, 1.43×, and 2.29× for the 2g.10gb, 3g.20gb, and 4g.20gb profiles, respectively, while experiencing a 0.6× decrease in the 7g.40gb profile. The final acceptance rate for GRMU is 22\% higher than that of the MCC policy. When compared to FF, a typical policy used in commercial solutions, GRMU offers a 39\% improvement.
	
	\subsubsection{Active Resources}

\begin{figure}[h]
		\centering
		\includegraphics[width=1\columnwidth]{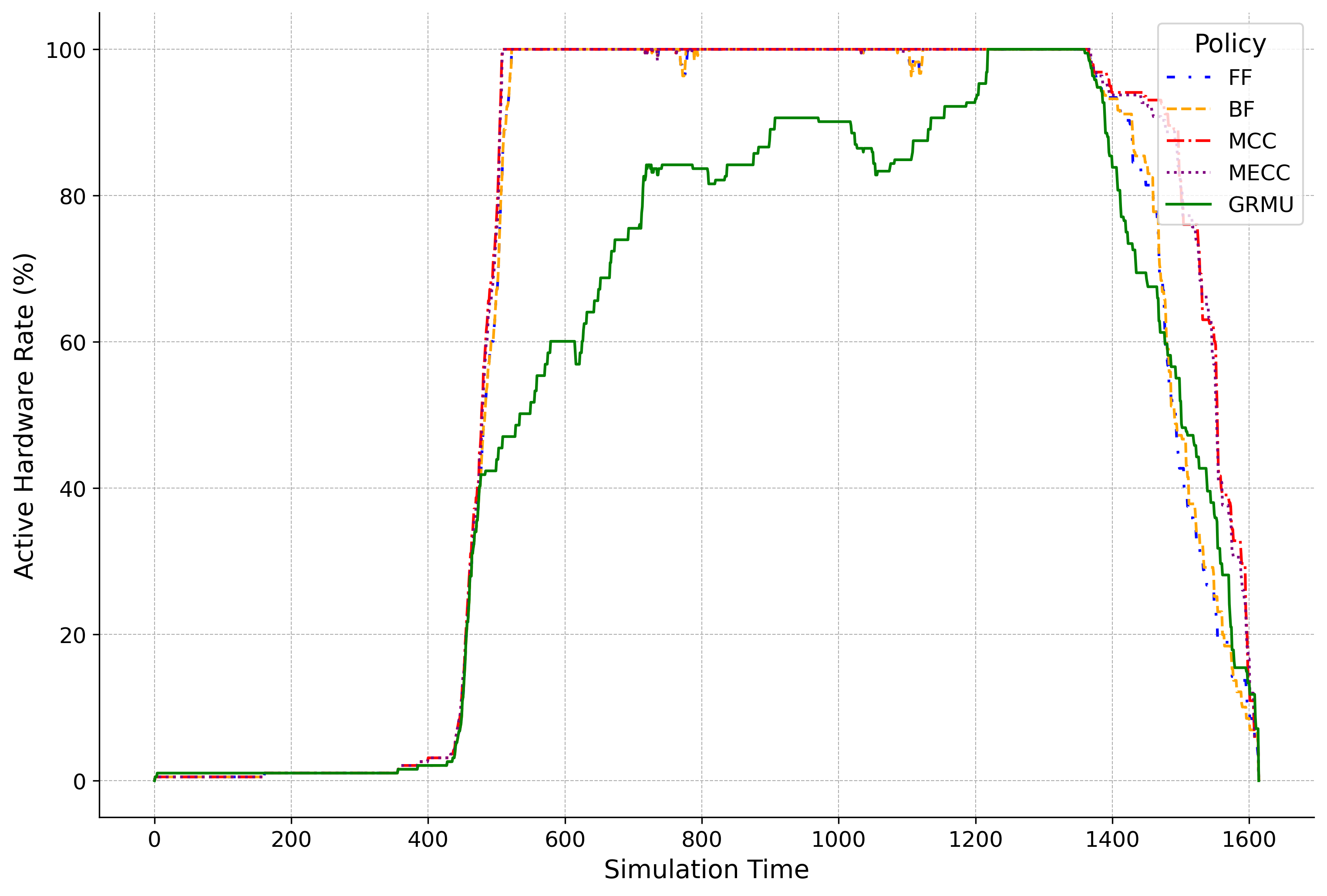}
		\caption{Active hardware rates per policy}
		\label{fig:active-resources-rates}
	\end{figure}

	Fig. \ref{fig:active-resources-rates} illustrates the trends in active hardware usage over simulation time for each policy. As shown, GRMU typically activates fewer resources during execution. For a more detailed comparison, Table \ref{tab:normalized_resource_utilization} presents the area under the curve for each policy in the chart. It is expected that FF and BF policies have a smaller cumulative area compared to MCC and MECC, as the latter distribute the load across more machines. Notably, GRMU outperforms the second-best policy, FF, by 17\%.

	\begin{table}[h!]
		\centering
		\begin{tabular}{|c|c|c|}
			\hline
			\textbf{Policy} & \textbf{Area Under the Curve} & \textbf{Normalized Value} \\ \hline
			FF             & 102,169.44                    & 0.9516                    \\ \hline
			BF             & 102,380.56                    & 0.9536                    \\ \hline
			MCC            & 107,363.19                    & 1.0000                    \\ \hline
			MECC           & 107,056.60                    & 0.9971                    \\ \hline
			GRMU           & 87,546.53                     & 0.8153                    \\ \hline
		\end{tabular}
		\caption{Cumulative active resource rate}
		\label{tab:normalized_resource_utilization}
	\end{table}
	
	\subsubsection{Migrations}
	
	The FF, BF, MCC, and MECC policies involve no migrations, while GRMU performs 37 migrations during execution. GRMU accepts 3,168 MIG-enabled VMs out of a total of 8,063. Therefore, the 37 migrations represent only 1\% of the total accepted guests, demonstrating its efficiency.

	\section{Conclusion}
	\label{s:conclusion}
	
	This article set out to tackle the challenges introduced by MIG technology in cloud data centers, particularly the fragmentation and inefficiency stemming from rigid placement rules. We formulated a multi-objective ILP model to capture the complexities of acceptance maximization, active hardware minimization, and reduced migration overhead. Building on this, we developed GRMU, a multi-stage placement framework designed to tackle MIG placement challenges using quota-based partitioning, defragmentation, and consolidation. Through evaluations using a real-world Alibaba GPU cluster trace, we showed that GRMU achieves improvements over state-of-the-art solutions. Specifically, GRMU increases the acceptance rate of incoming requests by 22\%, reduces active hardware usage by 17\%, and performs migrations on only 1\% of MIG-enabled VMs. This balance of higher acceptance and lower resource usage, all while incurring minimal migration overhead, underscores the efficacy of proposed solution. Future research could extend GRMU by incorporating more advanced prediction models for request patterns or integrating power-aware optimizations. As MIG becomes increasingly prevalent in modern cloud and HPC deployments, the insights and techniques described in this paper provide a foundation for enhanced GPU resource management in data centers.
	
	\bibliographystyle{elsarticle-num}
	\bibliography{bibliography}
	
\end{document}